\documentclass[aps,twocolumn,a4paper,showpacs]{revtex4}
\usepackage{graphicx}
\usepackage{amsmath}
\usepackage{amssymb}
\usepackage{enumerate, color}
\usepackage{subfigure}
\usepackage{tabularx}
\usepackage{epstopdf}
\newcommand{\be}{\begin{equation}}
\newcommand{\ee}{\end{equation}}
\newcommand{\ben}{\begin{eqnarray}}
\newcommand{\een}{\end{eqnarray}}
\newcommand{\bes}{\begin{subequations}}
\newcommand{\ees}{\end{subequations}}

\newcommand{\bb}{\bibitem}

\begin{document}
\title{How to build a compact brane}
\author{D. Bazeia}
\author{D.C. Moreira}
\affiliation{Departamento de F\'isica, Universidade Federal da Para\'iba, 58051-970, Jo\~ao Pessoa, PB, Brazil}
\begin{abstract}
This work deals with braneworld models in a five dimensional curved geometry with a single extra dimension of infinite extent.
The investigation introduces a new family of models, generated from a source scalar field that supports kinklike structures described through the presence of a real parameter, capable of controlling the thickness of the warp factor that describes the five dimensional geometry. The mechanism shows how to get a brane that engenders a compact profile.
\end{abstract}
\maketitle
\section{Introduction} 

Localized structures play important role in nonlinear science. In high energy physics, they may acquire different features and appear as, for instance, kinks, vortices and monopoles, in one, two and three spatial dimensions, respectively \cite{v,ms}. In the case of a single spatial dimension, kinks can be described by scalar field that self-interact through polynomial or nonpolynomial potentials. In the case of polynomial potentials, one usually considers fourth, sixth or higher-order polynomials, and in the case of nonpolynomial potentials, one has the sine-Gordon and the sinh-Gordon potentials as representative models.

In the two-dimensional spacetime, topological structures of the kinklike type can be used to study a diversity of problems in high energy physics, as we can see, for instance, in Refs.~\cite{a1,a2,a3,a4,a5}. Also, they are of current interest to source braneworld scenarios in $(4,1)$ spacetime dimensions, in a curved geometry with a single extra dimension of infinite extent \cite{c,rs,gw,d,fre,csaki,bc,gremm,bbn}.

In this work, we start dealing with real scalar field in $(1,1)$ spacetime dimensions, searching for kinklike structures and focusing mainly on the possibility to construct families of models, to be used to describe thick braneworld scenarios in $(4,1)$ spacetime dimensions. The work is motivated from the recent investigation on kinks and compactons, which offered the possibility to generate thick brane with hybrid profile \cite{blmm}. Here, however, we introduce a new mechanism, capable of inducing compact profile to the warp factor that describe the geometry of the brane.

The investigation concerns the study of scalar fields as source fields in a five-dimensional warped geometry, focusing on the possibility to navigate from models described by trigonometric interactions to other models. In order to get to the desired scenario, both in flat and in curved spacetime, we make use of the Jacobi elliptic functions, which are controlled by a single real parameter that connects trigonometric and hyperbolic functions. We start the investigation with the well-known standard $\phi^4$ theory with spontaneous symmetry breaking, which supports kinklike structure, and we make good use of the deformation procedure introduced in \cite{b1} to obtain the model described by elliptic functions, which we can solve analytically, using the prescription of Ref.~\cite{b1}. This route is of interest since it offers the possibility of obtaining analytical results, which we then use to investigate the braneworld scenario.

\section{Generalities}

We start with the Lagrange density
\be\label{ori}
\mathcal{L}=\frac{1}{2}\partial_{\mu}\phi\partial^{\mu}\phi-V(\phi),
\ee
where $\phi=\phi(x,t)$ represents the scalar field, $\mu=0,1$, and $V(\phi)$ is the potential, which is used to identify the model under investigation. Here we are working with dimensionless field and coordinates, for simplicity. The metric tensor is
$\eta_{\mu\nu}={\rm diag}\,(1,-1)$, and the energy-momentum tensor has the form
$T^{\mu\nu}=\partial^{\mu}\phi\partial^{\nu}\phi-\eta^{\mu\nu}\mathcal{L}$. The equation of motion for the scalar field is given by 
$\partial_{\mu}\partial^{\mu}\phi+{dV}/{d\phi}=0$; for static configuration we get
\begin{equation}
\frac{d^2\phi}{dx^2}=\frac{dV}{d\phi}.
\end{equation} 

We suppose that the potential $V(\phi)$ can be written as
\begin{equation}
V(\phi)=\frac{1}{2}w_{\phi}^2,
\end{equation}
where $w_{\phi}={dw}/{d\phi}$, with $w=w(\phi)$. If we write the potential in the form above, we get to the first-order equation
\begin{equation}
\frac{d\phi}{dx} =w_{\phi},
\end{equation}
which solves the equation of motion. The energy then becomes ${E}\!=\!|w\left(\phi (\infty)\right)\!-\!w\left(\phi \left(-\infty)\right)\right)|$, and 
the model can be seen as the bosonic portion of a supersymmetric model, with $w=w(\phi)$ known as the superpotential.

The strategy we shall follow in the current work is to use the deformation procedure put forward in \cite{b1}. It allows to introduce new models from known models: we choose the model \eqref{ori} and a specific potential $V(\phi)$ which we know to solve, to be the original model. We then introduce another field, $\chi$, and the function $f(\chi)$, such that the new potential $U(\chi)$ is given by
\begin{equation}
U(\chi)=\frac{V\left(\phi\rightarrow f(\chi)\right)}{f_{\chi}^2},
\end{equation}
where $f_\chi=df/d\chi$. The new theory is governed by the Lagrange density 
\begin{equation}\label{defmod}
{\mathcal{L}}(\chi,\partial_{\mu}\chi)=\frac{1}{2}\partial_{\mu}\chi\partial^{\mu}\chi-U(\chi),
\end{equation}
and we have, for static field $\chi=\chi(x)$
\begin{equation}\label{eq1}
\frac{d\chi}{dx}= {W}_{\chi},
\end{equation}
where ${W}(\chi)$  is the superpotential of the new model. It is such that
\be\label{W}
W_\chi=\frac{w_\phi(\phi\to f(\chi))}{f_\chi}.
\ee
In this case, the energy of the static field configuration is
\begin{equation}
{E}=|W\left(\chi (\infty)\right)-{W}\left(\chi(-\infty)\right)|.
\end{equation}

\begin{figure}[t]
\centerline{\includegraphics[height=14em]{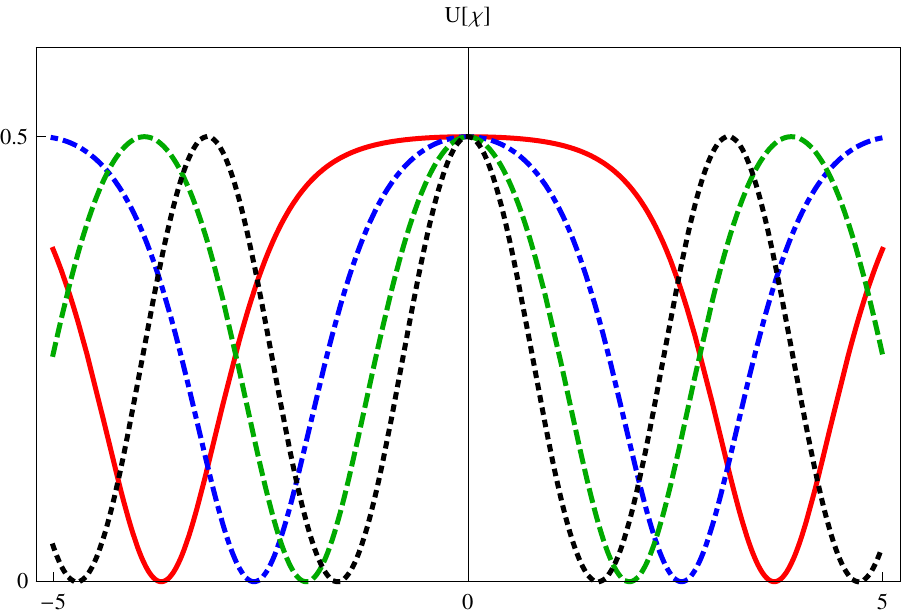}}
\caption{Plots of the potential \eqref{pot1}, for  $\lambda=0$ (black, dotted line) $0.6$ (green, dashed line), $0.9$ (blue, dot-dashed line) and $0.99$ (solid, red line)}\label{fig1}
\end{figure}

As it was shown in \cite{b1}, if $\phi(x)$ is static solution of the original, $\phi$-field model, then the solution of the new $\chi$-field model is given by 
\be\label{sol}
\chi(x)=f^{-1}(\phi(x)).
\ee
Another important issue is that the new model is described by the superpotential that appears from \eqref{W}, so the kinklike structures also solve the first-order differential equation \eqref{eq1}. As one knows, the solution \eqref{sol} of the model \eqref{defmod} is linearly stable, with the stability potential having the form
\be\label{stapot1}
U(x)= W^2_{\chi\chi}+ W_\chi W_{\chi\chi\chi}.
\ee
Here, $\chi=\chi(x)$ is the solution \eqref{sol} of the first-order Eq.~\eqref{eq1}, for $W_\chi$ given by Eq.~\eqref{W}. This potential entraps the zero mode, $\eta_0(x)$, which is proportional to the derivative of the solution itself ($\eta_0(x)\approx d\chi/dx$) and other bound states, depending of the specific form of the model under investigation. This first-order framework help us to find analytical solutions in the curved five-dimensional spacetime which we are interested in, to construct the braneworld scenario.

\section{Family of models}

An important model that supports kinklike solutions is the well-known $\lambda\phi^4$ model, described by the potential
\begin{equation}\label{phi4}
V(\phi)=\frac{1}{2}\left(1-\phi^2\right)^2.
\end{equation}
This potential has the $Z_2$ symmetry and has minima at $\bar{\phi}_{\pm}=\pm1$. The kinklike solution is
\begin{equation}  
\phi(x)=\tanh(x).
\end{equation}

\begin{figure}[t]
\centerline{\includegraphics[height=14em]{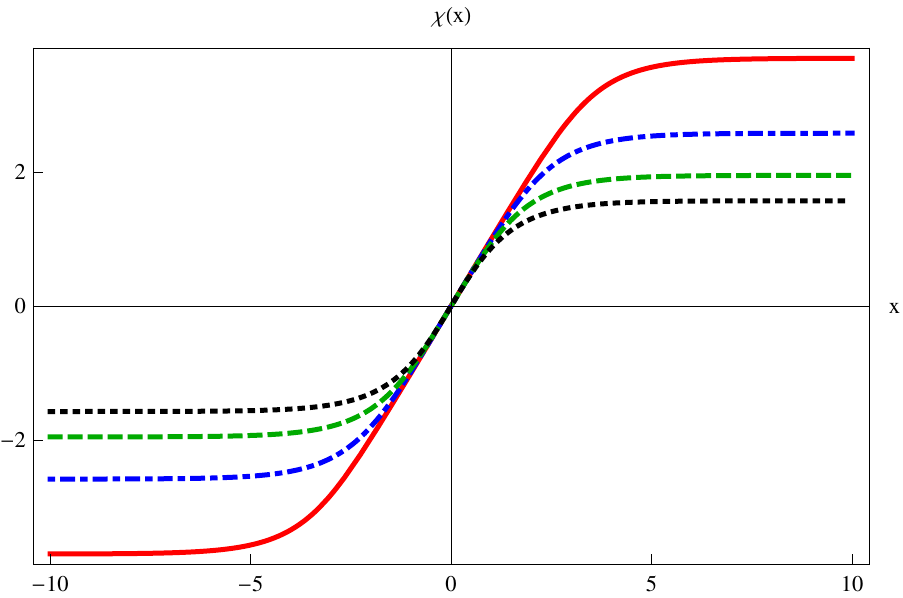}}
\caption{Plots of the solution \eqref{sol1} for some values of $\lambda$, as in Fig.~1.}\label{fig2}
\end{figure}

We use this model and the deformation function given by the Jacobi elliptic sine
\begin{equation}\label{d1}
f_\lambda(\chi)=\text{sn}(\chi, \lambda),
\end{equation}
where $\lambda$ is a real parameter, $\lambda\in[0,1]$; it is the modulus of the elliptic functions. We first recall that
\begin{eqnarray}
\text{cn}^2(\chi, \lambda)+\text{sn}^2(\chi, \lambda)&=&1,\\
\text{dn}^2(\chi, \lambda)+\lambda\text{sn}^2(\chi, \lambda)&=&1.
\end{eqnarray}
The elliptic functions are interesting since they lead to both trigonometric and hyperbolic functions; for $\lambda=0$ we have the usual trigonometric functions, and for $\lambda=1$ we get to the hyperbolic functions: in fact, $\text{sn}(\chi,0)=\sin(\chi)$, $\text{cn}(\chi,0)=\cos(\chi)$, and $\text{sn}(\chi, 1)=\text{tanh}(\chi)$ and $\text{cn}(\chi, 1)=\text{dn}(\chi, 1)=\text{sech}(\chi)$.
The functions $\text{sn}(\chi, \lambda)$ and $\text{cn}(\chi, \lambda)$ have period $4K$, while $\text{dn}(\chi, \lambda)$ has a smaller period, $2K$, with
\begin{equation}
K=\int_0^1 \frac{dt}{\sqrt{\left(1-t^2\right)\left(1-\lambda t^2\right)}}.
\end{equation}
Thus, if we use the function \eqref{d1} we get the potential
\begin{equation}\label{pot1}
\text{U}(\chi, \lambda)=\frac{1}{2}\text{cd}^2(\chi, \lambda),
\end{equation}
where $\text{cd}(\chi,\lambda)=\text{cn}(\chi, \lambda)/\text{dn}(\chi, \lambda)$. Also, 
\begin{equation}
\text{W}=-\frac{1}{\sqrt{\lambda}}{\ln \left(\frac{1-\sqrt{\lambda}\, \text{sn}(\chi , \lambda)}{\text{dn}(\chi , \lambda)}\right)},
\end{equation}
which leads to 
\be 
W_\chi=\text{cd}(\chi, \lambda)=\frac{\text{cn}(\chi, \lambda)}{\text{dn}(\chi, \lambda)}.
\ee

In Fig.~\ref{fig1} we depict the potential (\ref{pot1}). We see that the minima separate from each other as $\lambda$ increases from
$\lambda=0$ to $\lambda=1$. We see that for $\lambda=0$, the potential is of the sine-Gordon type, and in the limit $\lambda\to1$,
it becomes constant, $U\to1/2$. This is the behavior we want to use to build the braneworld scenario below. From the first-order equation one finds
\begin{equation}\label{1eq}
\chi'= \text{cd}\left(\chi, \lambda\right).
\end{equation}
The solution for the above equation is given by
\begin{equation}\label{sol1}
\chi(x, \lambda)=\text{sn}^{-1}\left(\text{tanh}(x), \lambda\right),
\end{equation}
which is depicted in Fig.~\ref{fig2} for some values of $\lambda$. The solution \eqref{sol1} is obtained from the deformation procedure \cite{b1}, and can be directly checked to be solution of \eqref{1eq}; see Eq.~\eqref{sol}.

\begin{figure}[t]
\centerline{\includegraphics[height=14em]{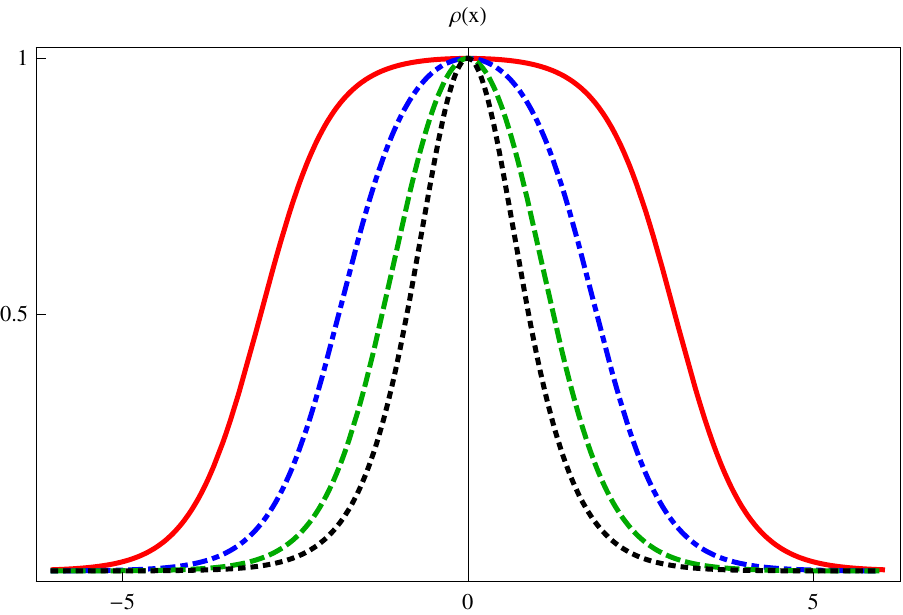}}
\caption{The energy density \eqref{ened} of the kinklike solution \eqref{sol1}, depicted for some values of $\lambda$, as in Fig.\ref{fig1}.}\label{fig3}
\end{figure}

To understand the behavior of the solution in the limit $\lambda\rightarrow 1$, we calculate its energy.
The energy density $T_{00}=\rho(x)$ has the form
\begin{equation}
\rho(x)=\frac{1}{2}\chi'^2+U(\chi)=\text{cd}^2(\chi, \lambda)
\end{equation}
or, explicitly in terms of $x$,
\begin{equation}\label{ened}
\rho(x)=\frac{1-\tanh^2( x)}{1-\lambda\tanh^2(x)}
\end{equation}
The total energy is written as
\begin{equation}\label{ene}
E(\lambda	)=|\Delta W|=\frac{1}{\sqrt{\lambda}}\ln\left(\frac{1+\sqrt{\lambda}}{1-\sqrt{\lambda}}\right).
\end{equation}
It starts at $E_0=2$, in the limit $\lambda\to0$, and diverges in the limit $\lambda\to1$. The issue here is that in the limit $\lambda\to1$, the potential becomes constant. As a consequence, the solution becomes a straight line, with constant derivative, leading the energy density to a constant value too, and making the energy to diverge as $\lambda$ approaches the unit value, as we illustrate in Figs.~\ref{fig1}, \ref{fig2} and \ref{fig3}.

The divergence found above makes the limit $\lambda\to1$ singular in $(1,1)$ spacetime dimensions. However, we identify that the problem is in the potential, which becomes constant in the limit $\lambda\to1$. Thus, if we go to higher dimensions and add the model into a given, curved geometry, the potential in the limit $\lambda\to1$ may describe a cosmological constant, circumventing the problem found in the $(1,1)$ spacetime. 

\subsection{Braneworld scenario}

\begin{figure}[t]
\centerline{\includegraphics[height=14em]{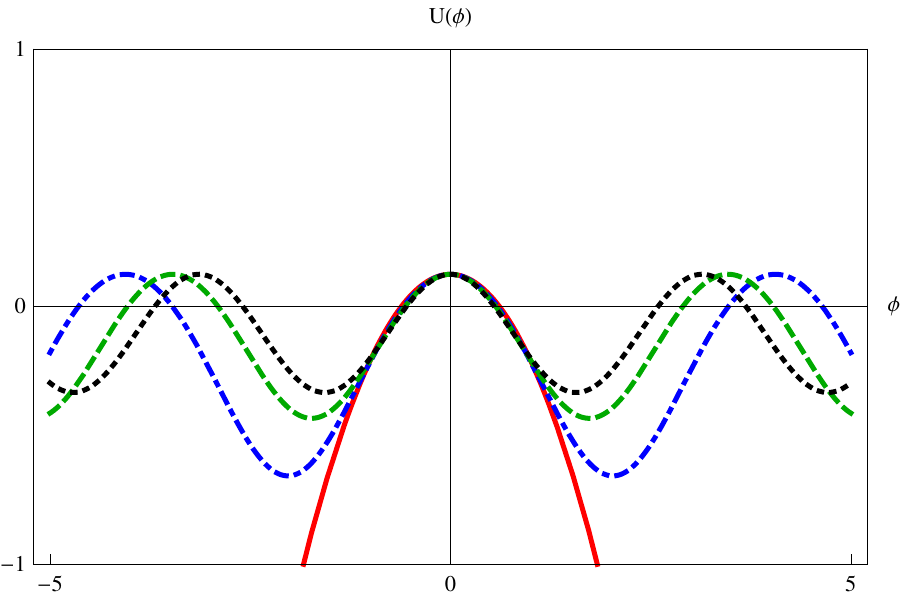}}
\caption{The potential \eqref{potb1} of the source scalar field in the braneworld model, depicted for $\lambda=0$ (black, dotted line), $\lambda=1/3$ (green, dashed line), $2/3$ (blue, dot-dashed line) and $\lambda=1$ (red, solid line)}\label{fig4}
\end{figure}
\begin{figure}[t]
\centerline{\includegraphics[height=14em]{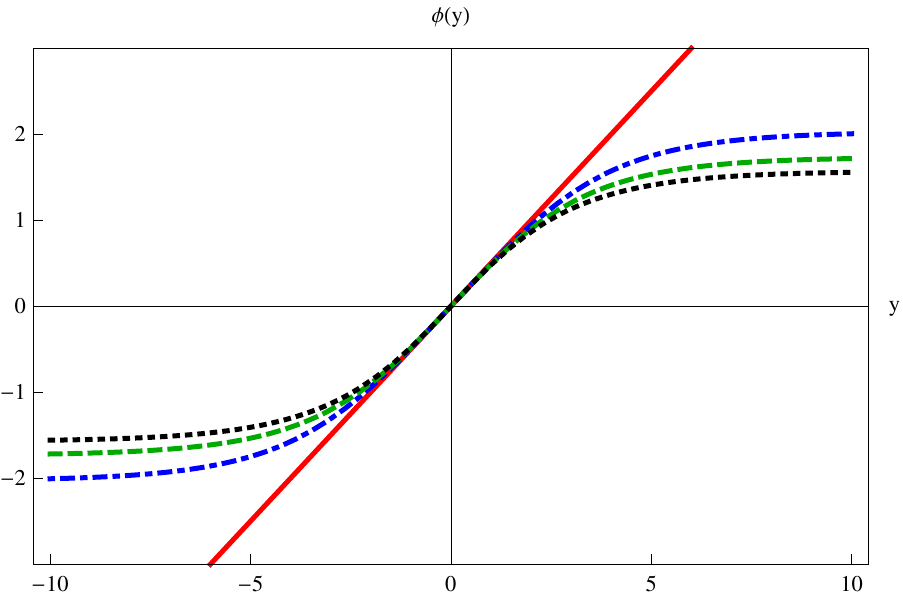}}
\caption{The scalar field solution \eqref{solb2} of the braneworld model \eqref{potb1}, depicted for $\lambda$ as in Fig.~\ref{fig4}}\label{fig5}
\end{figure}

With the above motivation in mind, we now turn attention to the case of a braneworld model described by a five dimensional warped geometry, with a single extra dimension of infinite extent \cite{rs,gw,d,fre,csaki,bc,gremm,bbn}. We consider the Einstein-Hilbert action which reads 
\begin{equation}
S=\int d^5x\sqrt{|g|}\left(\frac{1}{4}R+\mathcal{L}\right).
\end{equation}
Here we are using $4\pi G_5=1$ and the Lagrange density of the source field is given by
\begin{equation}
\mathcal{L}(\phi,\partial_{a}\phi)=\frac{1}{2}g_{ab}\partial^{a}\phi\partial^{b}\phi-U(\phi).
\end{equation}
The metric is
\begin{equation}
ds^2_5=g_{ab}dx^a dx^b=e^{2A(y)}ds^2_4-dy^2
\end{equation}
where $ds^2_4=\eta_{\mu\nu}dx^{\mu}dx^{\nu}$ and $y$ describes the extra spatial dimension. The Einstein equation has the general form
\begin{equation}
G_{ab}=2T_{ab},
\end{equation}
As usual, we take $A=A(y)$ and $\phi=\phi(y)$, and so we get
\bes\label{29}\begin{eqnarray}
 6A'^2&=&\phi'^2-2U,\\
3A''+6A'^2&=&-\phi^{\prime 2}-2 U.
\end{eqnarray}\ees
Here prime stands for derivative with respect to $y$. We use \eqref{29} to get
\begin{equation}
A''=-\frac{2}{3}\phi'^2. 
\end{equation}
We then introduce the first-order equations 
\bes\label{foequa}\ben
\phi'&=&\frac12 W_{\phi},\label{field}
\\
A'&=&-\frac{1}{3}\,W,
\een\ees
which solve the equations of motion if the potential is given by
\be\label{poten}
U(\phi)=\frac18\,W_\phi^2-\frac13\, W^2.
\ee

An interesting feature that appears from the above braneworld scenario is that the total energy vanishes. To see this, we note that the energy density, which is given by
\be
\rho(y)= e^{2 A}\left(\frac12\phi^\prime+U(\phi)\right),
\ee 
can be written in the form, with the use of Eqs.~\eqref{foequa} and \eqref{poten},
\be
\rho(y)=\frac12\frac{d}{dy}\left(W\,e^{2A}\right),
\ee
so it is such that the total energy usually adds to zero, since the warp factor $\exp{(2A)}$ vanishes exponentially, asymptotically.

\subsection{Metric fluctuations}

Another general feature concerns stability of the gravity sector of the braneworld model. To show this, 
we redefine the $y-$coordinate to $dy^2=e^{2A(z)}dz^2$ and take the metric to the form
\begin{equation}
ds^2=e^{2A(z)}\left(\eta_{ab}+h_{ab}\right)dx^a dx^b.
\end{equation}
We then deal with conformal related metrics, i.e.,  $g_{ab}=e^{2A}\tilde{g}_{ab}$. Using transverse-traceless gauge one can write
${\tilde G}=-\frac12\partial_c\partial^c h_{ab}$, which leads to the result
\begin{eqnarray}
\nonumber G_{ab}^{(1)}&=&-\frac{1}{2}\partial_c\partial^c h_{ab}+3\Bigl[\partial_a A\partial_b A-\partial_a \partial_b A+\\
&~&+\frac{1}{2}A'h'_{ab}+\tilde{g}_{ab}\left(\partial_c \partial^c A+\partial_c A\partial^c A\right)\Bigr].
\end{eqnarray}
Thus, the $\mu\nu-$components of the linearized tensor are
\begin{equation}
G_{\mu\nu}^{(1)}=-\frac{1}{2}\partial_c\partial^c h_{\mu\nu}+\frac{3}{2}A'h'_{\mu\nu}-3\tilde{g}_{\mu\nu}\left( A''+ A'^2\right).
\end{equation}
Here prime means derivative with respect to $z$.

\begin{figure}[t!]
\centerline{\includegraphics[height=15em]{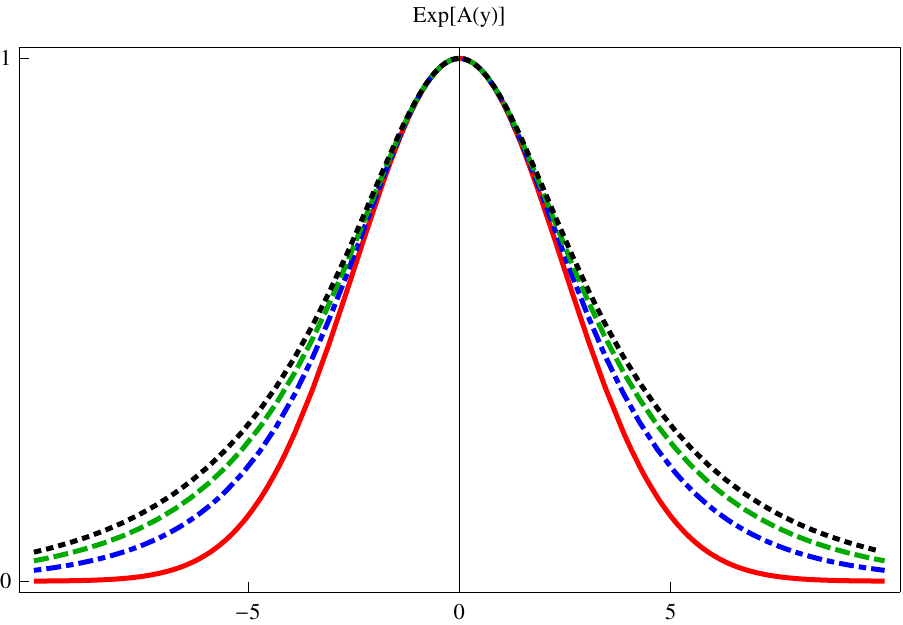}}
\caption{The warp factor, depicted for $\lambda$ as in Fig.~\ref{fig4}.}\label{fig6}
\end{figure}

\begin{figure}[t!]
\centerline{\includegraphics[height=14em]{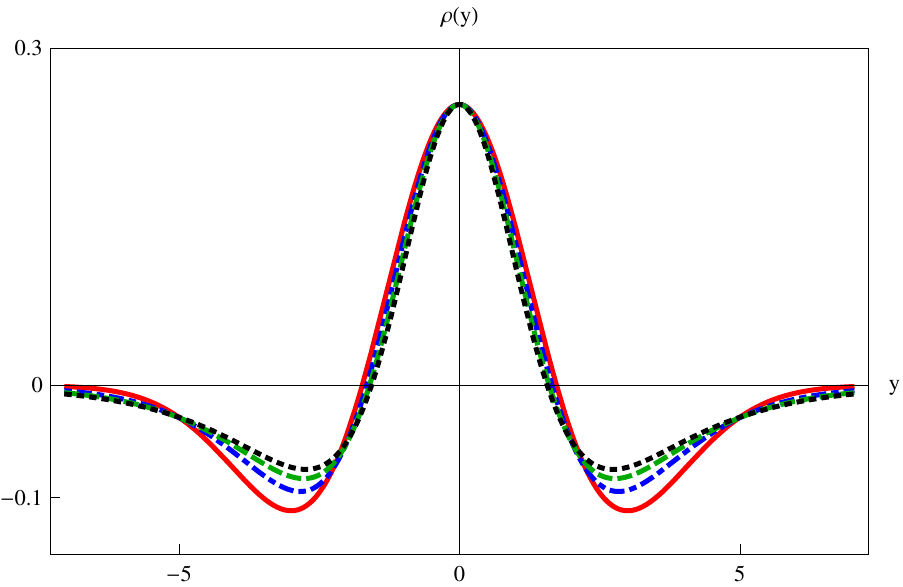}}
\caption{The energy density \eqref{ened2}, depicted for $\lambda$ as in Fig.~\ref{fig4}.}\label{fig7}
\end{figure}

\begin{figure}[t]
\centerline{\includegraphics[height=14em]{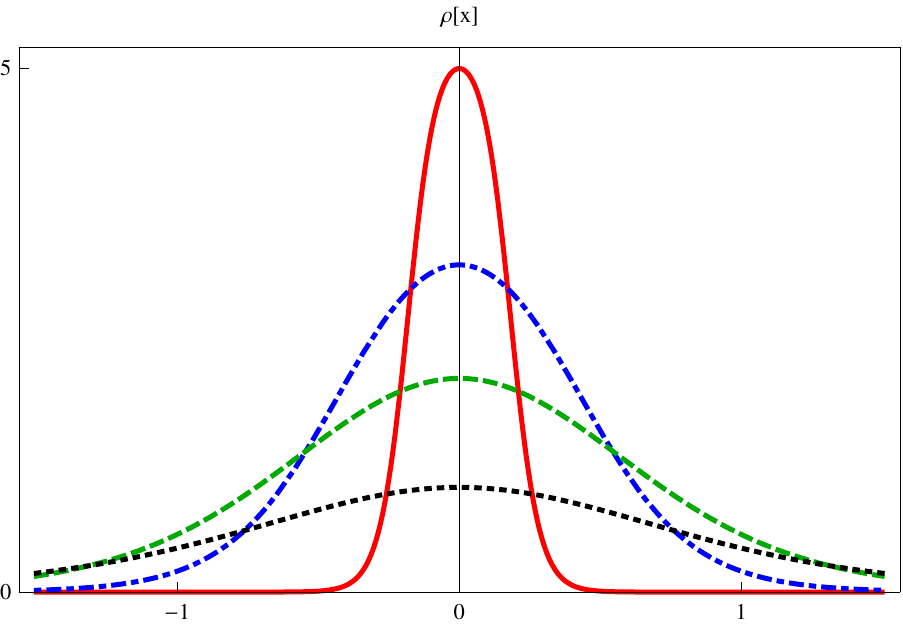}}
\caption{The energy density, depicted for $\lambda=0$ (black, dotted line), $\lambda=0.3$ (green, dashed line), $\lambda=0.6$ (blue, dot-dashed line) and $\lambda=0.9$ ($\rho/3$, red, solid line).}\label{fig8}
\end{figure}
\begin{figure}[t]
\centerline{\includegraphics[height=14em]{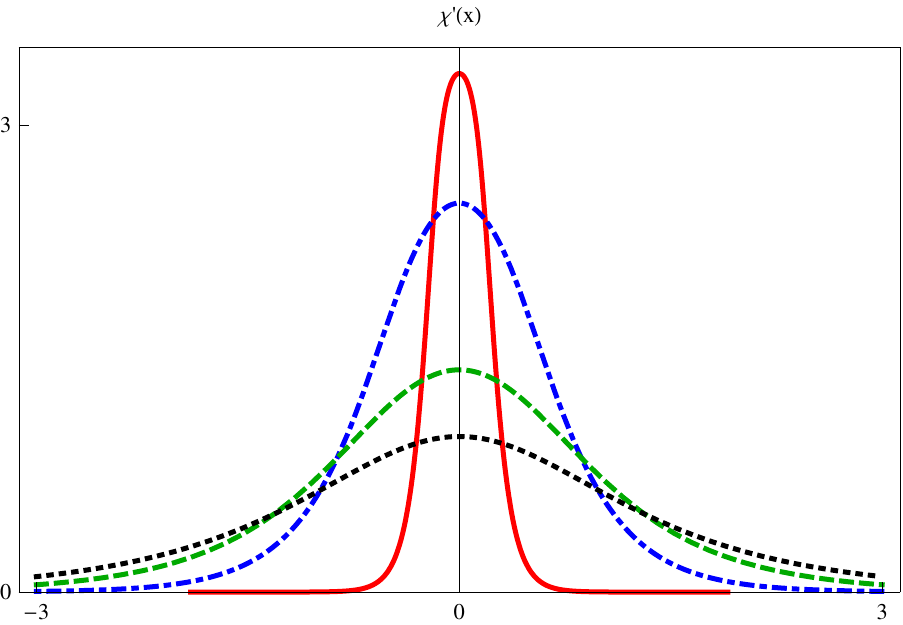}}
\caption{The zero mode, depicted for $\lambda=0$ (black, dotted line), $\lambda=0.3$ (green, dashed line), $\lambda=0.6$ ($\chi^\prime/2$, blue, dot-dashed line) and $\lambda=0.9$ ($\chi^\prime/20$, red, solid line).}\label{fig9}
\end{figure}

We can write the linearized energy-momentum tensor as $2T_{\mu\nu}^{(1)}=-3\tilde{g}_{\mu\nu}\left( A''+ A'^2\right)$.
We then use $G_{\mu\nu}^{(1)}=2T_{\mu\nu}^{(1)}$ to get
\begin{equation}
-\partial_c\partial^c h_{\mu\nu}+3A'h'_{\mu\nu}=0,
\end{equation}
which, under the transformation $H_{\mu\nu}=e^{-ipx}e^{3A/2}h_{\mu\nu}$, turns out to be
\begin{equation}
\left(\partial_z+\frac{3}{2}A'\right)\left(-\partial_z+\frac{3}{2}A'\right)H_{\mu\nu}=p^2 H_{\mu\nu}.
\end{equation}
This equation has the form $S^{\dagger}S\psi=p^2\psi$, and shows that the Hermitian operator $S^{\dagger}S$ is non-negative, as it should be to describe linearly stable and consistent gravity sector. The stability potential is written in the form
\be\label{sp}
U(z)=\frac32 A^{\prime\prime}+ \frac94 A^{\prime 2}.
\ee
It has the usual profile and may support the zero mode (the graviton) as a stable state, bounded to the brane.   
\subsection{Models}

Inspired from the study on the flat spacetime, we consider the function
\begin{equation}
W=-\frac{1}{\sqrt{\lambda}}{\ln \left(\frac{1-\sqrt{\lambda} \,\text{sn}(\phi, \lambda)}{\text{dn}(\phi, \lambda)}\right)},
\end{equation}
such that
\be\label{wp}
W_{\phi}=\frac{\text{cn}(\phi, \lambda)}{\text{dn}(\phi, \lambda)}.
\ee
The potential then becomes
\be\label{potb1}
U(\phi)=\frac{\text{cn}(\phi, \lambda)^2}{8 \text{dn}(\phi, \lambda)^2}-\frac1{3\lambda}{\ln^2\!\left(\frac{1-\sqrt{\lambda}\, \text{sn}(\phi, \lambda)}{\text{dn}(\phi, \lambda)}\right)}.\;\;\;
\ee
It is depicted in Fig.~\ref{fig4} for some values of $\lambda$.
The field solution is
\begin{equation}\label{solb2}
\phi(y)=\text{sn}^{-1}(\tanh ({y}/{2}), \lambda),
\end{equation}
as given by Eq.~\eqref{sol}, obtained from the deformation procedure. It is depicted in Fig.~\ref{fig5} for some values of $\lambda$.

Note that in the limit $\lambda\rightarrow 0$, we have $U(\phi)=\frac{1}{48} (11 \cos (2 \phi )-5)$, and for $\lambda\rightarrow 1$, one finds $U(\phi)=\frac{1}{8}-\frac{1}{3} \ln ^2(\cosh (\phi )-\sinh (\phi ))$. The case $\lambda=0$ leads to a particular case of the sine-Gordon model studied in \cite{gremm}, for $b=2/3$ and $c=1/2$. 

We have been unable to find an analytical expression for the warp factor, so we depict the numerical solution for some values of $\lambda$ in Fig.~\ref{fig6}.
Also, the energy density is given by
\begin{equation}\label{ened2}
\rho(y)=e^{2A(y)}\left(\frac{\text{cn}(\phi, \lambda)^2}{4 \text{dn}(\phi, \lambda)^2}-\frac{\ln ^2\left(\frac{1-\sqrt{\lambda} \text{sn}(\phi, \lambda)}{\text{dn}(\phi, \lambda)}\right)}{3 \lambda}\right).
\end{equation}
It is depicted in Fig.~\ref{fig7} for some values of $\lambda$.

One sees from Fig.~\ref{fig5} that if one increases the value of $\lambda$, the amplitude of the solution also increases, but it makes no important change to its derivative near the origin. The effect of this in the geometry is to slowly narrow the warp factor, as it appears in Fig.~\ref{fig6}.

\begin{figure}[t]
\centerline{\includegraphics[height=14em]{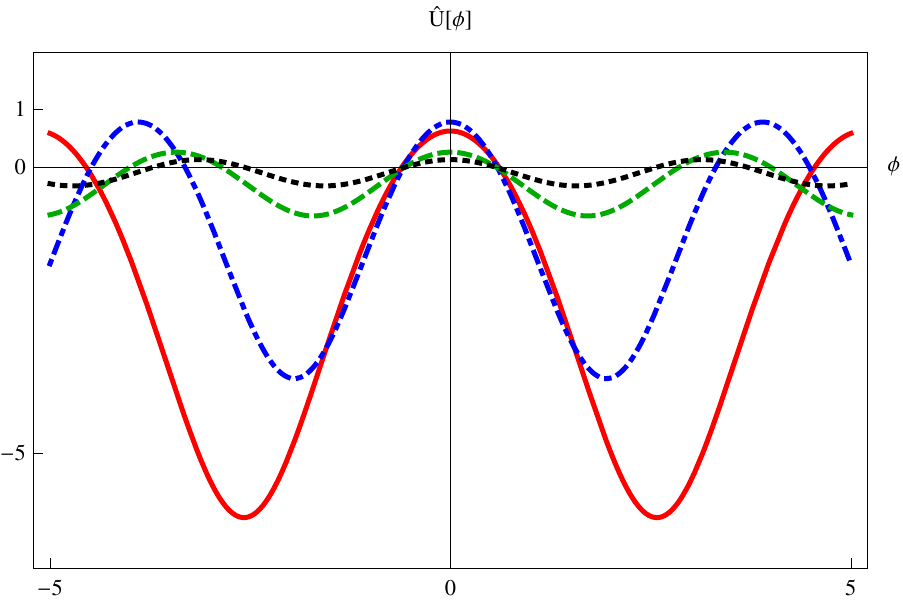}}
\caption{The potential of the new model, depicted for $\lambda=0$ (black, dotted line), $\lambda=0.3$ (green, dashed line), $\lambda=0.6$ (blue, dot-dashed line) and $\lambda=0.9$ (${\tilde{U}}$/20, red, solid line).}\label{fig10}
\end{figure}
\begin{figure}[t]
\centerline{\includegraphics[height=14em]{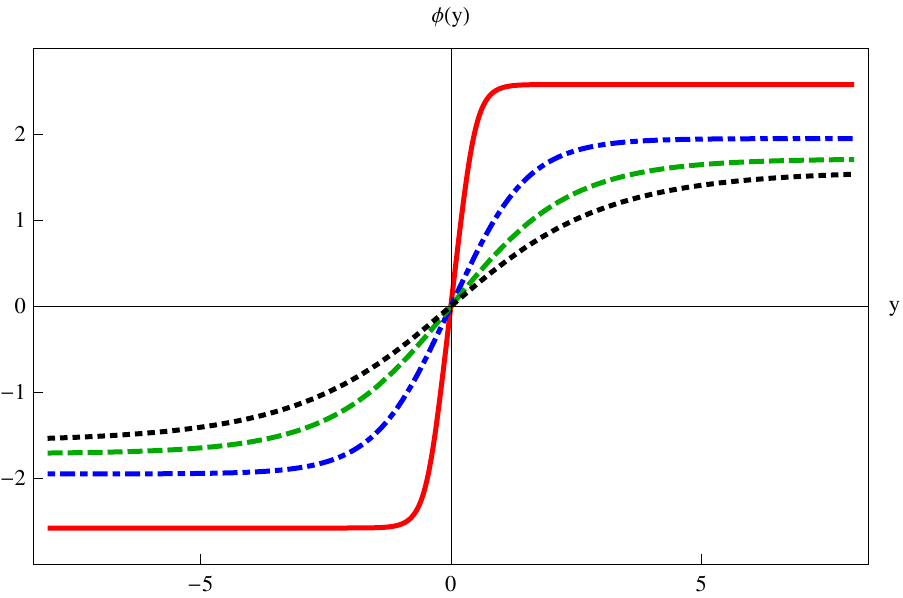}}
\caption{The kinklike solution \eqref{solc}, depicted for $\lambda$ as in Fig.~\ref{fig10}.}\label{fig11}
\end{figure}

The above results teach us that if we want to make the warp factor more localized, we have to increase the derivative of the scalar field solution near the origin. However, from the first-order equation \eqref{field} we see that this can be implemented with the increasing of the value of $W_\phi$, which is given by \eqref{wp}. We then make the following change: we rewrite the superpotential of the deformed model to the form
\begin{equation}\label{wc}
W=-\frac{1}{1-\lambda}\frac{1}{\sqrt{\lambda}}{\ln \left(\frac{1-\sqrt{\lambda} \text{sn}(\phi, \lambda)}{\text{dn}(\phi, \lambda)}\right)},
\end{equation}
such that
\be
W_\phi=\frac{1}{1-\lambda}\frac{\text{cn}(\phi, \lambda)}{\text{dn}(\phi, \lambda)}.
\ee
This modification requires that we exclude the case $\lambda=1$, so we have to take $\lambda\in[0,1)$ from now on.
The new potential becomes $\tilde{U}(\phi)=U(\phi)/(1-\lambda)^2$, where $U(\phi)$ is given by \eqref{potb1}, and the new solution is
\be \label{solc}
\phi(y)=\text{sn}^{-1}\left(\tanh ({y}/{2(1-\lambda)},\lambda\right).
\ee
We study this kinklike solution in $(1,1)$ spacetime dimensions to see how it behaves as $\lambda$ increases in the interval $[0,1)$. We depict the energy density and the corresponding (non normalized) zero mode in Figs.~\ref{fig8} and \ref{fig9}, respectively. They both show very clearly the tendency to concentrate around the origin as $\lambda$ increases toward unit, so it may contribute to make the brane compact. 

\begin{figure}[t]
\centerline{\includegraphics[height=14em]{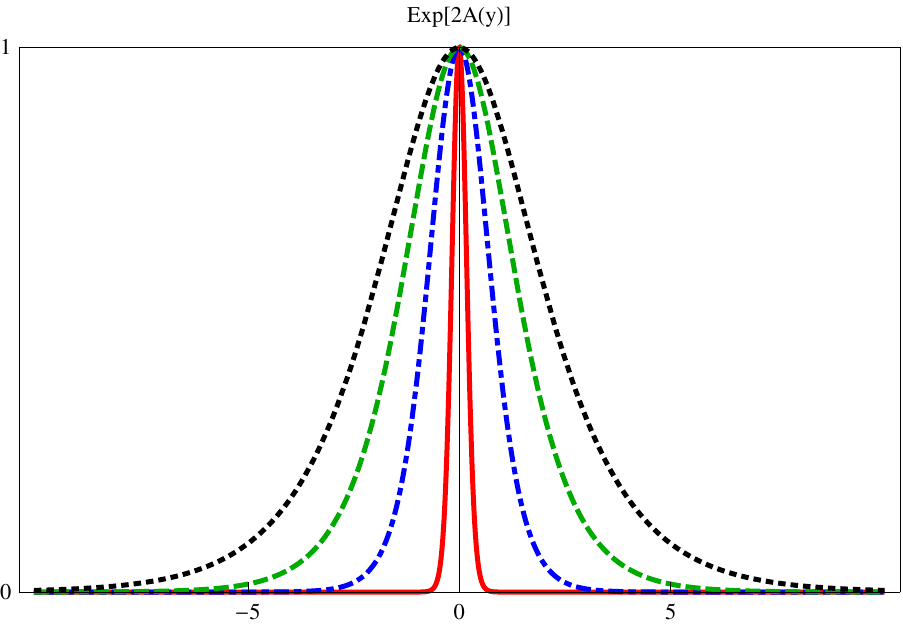}}
\caption{The warp factor, depicted for $\lambda$ as in Fig.~\ref{fig10}.}\label{fig12}
\end{figure}

\begin{figure}[t]
\bigskip
\centerline{\includegraphics[height=14em]{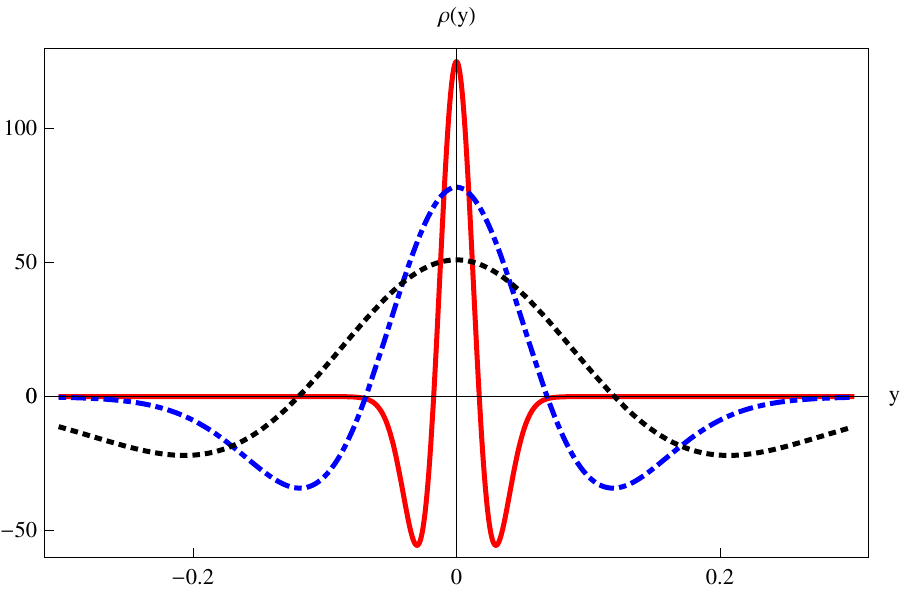}}
\caption{The energy density $\rho$, depicted for $\lambda=0.93$ (black, dotted line), $\lambda=0.96$ (blue, dot-dashed line, $\rho$/2) and $\lambda=0.99$ ($\rho/20$, red, solid line).}\label{fig13}
\end{figure}

We then turn attention to the braneworld scenario. The new potential ${\tilde U}(\phi)$ is depicted in Fig.~\ref{fig10} and the solution \eqref{solc} in Fig.~\ref{fig11}. We see that the solution becomes thinner and thinner, and the amplitude increases as $\lambda$ increases in the interval $\lambda\in[0,1)$.  The new warp factor is depicted in Fig.~\ref{fig12} and in Fig.~\ref{fig13} we plot the energy density $\rho$ for some values of $\lambda$ very close to unit. It is interesting to note that in the limit $\lambda\to1$, the kinklike solution behaves as the singular tachyon kink introduced sometime ago; see, e.g., Ref.~\cite{sen} and references therein. However, in the case of $\lambda\approx 1$ the model describes braneworld configuration which engenders a regular behavior, which is very well localized around the origin; as $\lambda$ increases toward unit, the warp function becomes more and more localized, vanishing exponentially very strongly, and smoothly controlling the energy density such that the total energy vanishes, as we illustrate in Fig.~{\ref{fig13}}. 

The behavior of the warp factor in Fig.~\ref{fig12} shows that its width decreases significantly as $\lambda$ increases toward unit, making the extra dimension compact. The procedure presented above offers a new braneworld scenario, which has nothing to do with the mechanisms explored before in \cite{bc} and more recently in \cite{blmm}. 

\begin{figure}[h]
\bigskip
\centerline{\includegraphics[height=14em]{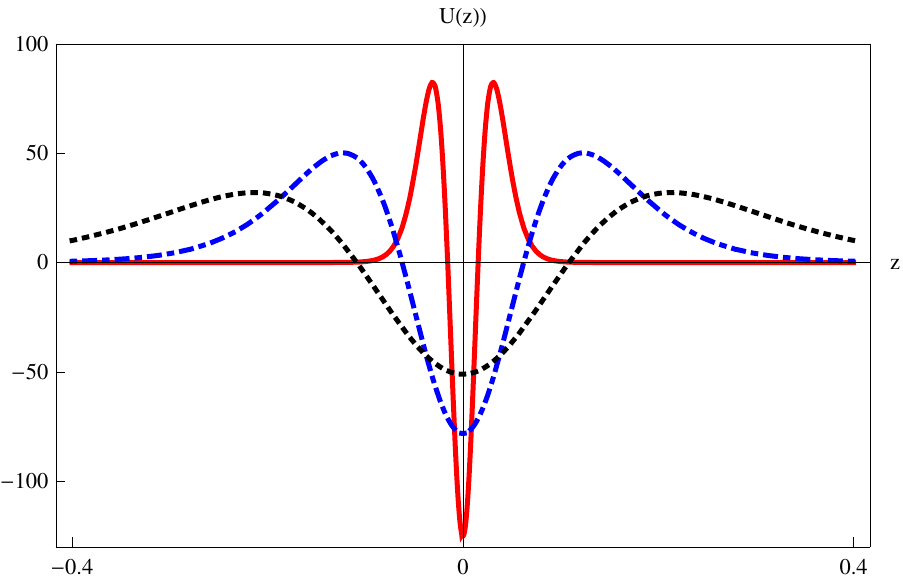}}
\caption{The stability potential \eqref{sp} for the model \eqref{wc}, depicted for $\lambda$ as in Fig.~\ref{fig13}, with the same factors $1$, $1/2$, and $1/20$, respectively.}\label{fig14}
\end{figure}

To see if the gravity sector is robust, we focus on the stability potential \eqref{sp} for the model under investigation. Here we follow \cite{sta} to study its profile, to see that it behaves smoothly even for $\lambda$ very close to unit. We then depict it in Fig.~\ref{fig14} for some values of $\lambda$ close to unit. We note that the potential narrows as $\lambda$ increases toward unit, but it keeps the zero mode bounded to the brane.

As $\lambda$ increases toward unit, the warp factor of the brane becomes thinner and thinner, but it is well different from the thin brane profile originally introduced in \cite{rs}. We have calculated the Kretschmann scalar to get
\be  
R^{abcd}R_{abcd}=40A^{\prime4}+16 A^{\prime\prime2}+32A^{\prime2}A^{\prime\prime}.
\ee
We checked that it behaves smoothly for $\lambda$ in the interval $[0,1)$. It increases for increasing $\lambda$, and diverges at the origin $y=0$ for $\lambda=1$, as expected. We then conclude that the brane has a regular behavior as $\lambda$ varies in the interval $\in[0,1)$, and tends to become compact as $\lambda$ approaches unit.


\section{Ending comments}

In this work we studied braneworld models described by a single real scalar field in a warped geometry with a single extra spatial dimension of infinite extent. We used the deformation procedure suggested in \cite{b1} to build a scalar field model which is described by Jacobi elliptic functions, controlled by a single real parameter that acquires values in the interval $\lambda\in[0,1)$. The interesting result is the presence of a new braneworld scenario, which describes a regular zero energy brane configuration with the warp function more and more localized around its center, as $\lambda$ increases toward unit.

The braneworld model is controlled by a single real parameter $\lambda\in[0,1)$, and its construction has nothing to do with the investigations described before in \cite{blmm}. As we can see from the energy density and the stability potential depicted in Figs.~\ref{fig13} and \ref{fig14}, the brane behaves regularly for $\lambda$ very close to unit. The model that we proposed leads to a new braneworld scenario, well different from the thin or thick brane profile introduced sometime ago.

The proposed model is described by a parameter that controls the thickness of the brane in a very nice way, and may have applications of current interest to particle phenomenology. An interesting line of investigation would be to study the presence of fermions and other fields, to verify how they can be entrapped inside the brane, controlled
by $\lambda$.

\acknowledgements{We thank the Brazilian agencies CAPES and CNPq for financial support.}



\begin{thebibliography}{99}
\bibitem{v}A. Vilenkin and E.P.S. Shellard, {\it Cosmic strings and other topological defects} (Cambridge UP, Cambridge, UK, 1994).
\bibitem{ms}N.S. Manton and P. Sutcliffe, {\it Topological Solitons} (Cambridge UP, Cambridge, UK, 2004).
\bb{a1}G. Basar and G.V. Dunne, Phys. Rev. Lett. {\bf100}, 200404 (2008).
\bb{a2}S. Dutta, D. A. Steer, and T. Vachaspati, Phys. Rev. Lett. {\bf101}, 121601 (2008).
\bb{a3}A. Alonso-Izquierdo, M.A. Gonzalez Leon, and J. Mateos Guilarte, Phys. Rev. Lett. {\bf101}, 131602 (2008).
\bb{a4}T. Romanczukiewicz and Ya. Shnir, Phys. Rev. Lett. {\bf105}, 081601 (2010)
\bb{a5}P. Dorey, K. Mersh, T. Romanczukiewicz, and Ya. Shnir, Phys. Rev. Lett. {\bf107}, 091602 (2011).
\bb{c}M. Cvetic, S. Griffies, S.-J. Rey, Nucl. Phys. B {\bf381}, 301 (1992).
\bibitem{rs}L. Randall and R. Sundrum, Phys. Rev. Lett. \textbf{83}, 4690 (1999).
\bibitem{gw}W.D. Goldberger and M.B. Wise, Phys. Rev. Lett. {\bf83}, 4922 (1999).
\bb{d}G. Dvali, G. Gabadadze, M. Porrati, Phys. Lett. B {\bf485}, 208 (2000).
\bibitem{fre}O. DeWolfe, D.Z. Freedman, S.S. Gubser and A. Karch, Phys. Rev. D {\bf62}, 046008 (2000).
\bb{csaki}C. Csaki, J. Erlich, T.J. Hollowood, and Y. Shirman, Nucl. Phys. B {\bf581}, 309 (2000).
\bibitem{gremm} M. Gremm, Phys. Lett. B \textbf{478}, 434 (2000).
\bb{bc}F.A. Brito, M. Cvetic, and S. Yoon, Phys. Rev. D {\bf64}, 064021 (2001).
\bb{bbn}D. Bazeia, F.A. Brito, and J.R. Nascimento, Phys. Rev. D {\bf68}, 085007 (2003);
D. Bazeia, C. Furtado, and A.R. Gomes, JCAP {\bf0402}, 002 (2004).
\bb{blmm}D. Bazeia, L. Losano, M.A. Marques, and R. Menezes, Phys. Lett. B {\bf736}, 515 (2014). 
\bibitem{b1} D. Bazeia, L. Losano and J.C.M Malbouisson, Phys. Rev D  \textbf{66}, R101701 (2002).
\bibitem{sen}Ashoke Sen, Int. J. Mod. Phys. A {\bf20}, 5513 (2005).
\bb{sta}D. Bazeia, A.R. Gomes, and L. Losano, Int. J. Mod. Phys. A {\bf24}, 1135 (2007). 
\end{thebibliography}
\end{document}